\documentclass[sigconf]{acmart}
\AtBeginDocument{%
  }

\copyrightyear{2025}
\acmYear{2025}
\setcopyright{acmlicensed}
\acmConference[MM '25] {Proceedings of the 33rd ACM International Conference on Multimedia}{October 27--31, 2025}{Dublin, Ireland.}
\acmBooktitle{Proceedings of the 33rd ACM International Conference on Multimedia (MM '25), October 27--31, 2025, Dublin, Ireland}
\acmISBN{979-8-4007-2035-2/2025/10}
\acmDOI{10.1145/3746027.3754988}
\usepackage{multirow}
\usepackage[normalem]{ulem}
\useunder{\uline}{\ul}{}
\usepackage{bm}
\usepackage{fmtcount}
\usepackage{pifont}
\usepackage{graphicx}

\begin{document}

\title[Cross-Domain Collaborative General Speech Enhancement via Hierarchical Language Models]{From Continuous to Discrete: Cross-Domain Collaborative General Speech Enhancement via Hierarchical Language Models}
\author{Zhaoxi Mu}
\authornote{Work done during an internship at Tencent AI Lab.}
\orcid{0009-0004-7393-0425}
\affiliation{%
  \institution{Xi'an Jiaotong University}
  \city{Xi'an}
  \state{Shaanxi}
  \country{China}
}
\email{wsmzxxh@stu.xjtu.edu.cn}

\author{Rilin Chen}
\orcid{0009-0004-6873-1683}
\affiliation{%
  \institution{Tencent AI Lab}
  \city{Beijing}
  \country{China}
}
\email{rilinchen@tencent.com}

\author{Andong Li}
\orcid{0000-0003-4094-8448}
\affiliation{%
  \institution{Institute of Acoustics, Chinese Academy of Sciences}
  \city{Beijing}
  \country{China}
}
\email{liandong@mail.ioa.ac.cn}

\author{Meng Yu}
\orcid{0000-0002-0031-9156}
\affiliation{%
  \institution{Tencent AI Lab}
  \city{Bellevue}
  \state{WA}
  \country{USA}
}
\email{raymondmyu@global.tencent.com}

\author{Xinyu Yang}
\authornote{Corresponding author.}
\orcid{0000-0001-5117-4914}
\affiliation{%
  \institution{Xi'an Jiaotong University}
  \city{Xi'an}
  \state{Shaanxi}
  \country{China}
}
\email{yxyphd@mail.xjtu.edu.cn}

\author{Dong Yu}
\orcid{0000-0003-0520-6844}
\affiliation{%
  \institution{Tencent AI Lab}
  \city{Bellevue}
  \state{WA}
  \country{USA}
}
\email{dyu@global.tencent.com}

\renewcommand{\shortauthors}{Zhaoxi Mu et al.}

\begin{abstract}
    This paper introduces OmniGSE, a novel general speech enhancement (GSE) framework designed to mitigate the diverse distortions that speech signals encounter in real-world scenarios. These distortions include background noise, reverberation, bandwidth limitations, signal clipping, and network packet loss. Existing methods typically focus on optimizing for a single type of distortion, often struggling to effectively handle the simultaneous presence of multiple distortions in complex scenarios. OmniGSE bridges this gap by integrating the strengths of discriminative and generative approaches through a two-stage architecture that enables cross-domain collaborative optimization. In the first stage, continuous features are enhanced using a lightweight channel-split NAC-RoFormer. In the second stage, discrete tokens are generated to reconstruct high-quality speech through language models. Specifically, we designed a hierarchical language model structure consisting of a RootLM and multiple BranchLMs. The RootLM models general acoustic features across codebook layers, while the BranchLMs explicitly capture the progressive relationships between different codebook levels. Experimental results demonstrate that OmniGSE surpasses existing models across multiple benchmarks, particularly excelling in scenarios involving compound distortions. These findings underscore the framework's potential for robust and versatile speech enhancement in real-world applications.
\end{abstract}



\maketitle

\section{Introduction}
Speech enhancement (SE) aims to improve the quality and intelligibility of speech signals, with applications spanning communication systems, hearing aids, speech recognition, and real-time audio/video conferencing. In real-world environments, speech signals are often subjected to mixed distortions, including background noise, room reverberation, bandwidth limitations, signal clipping, and network packet loss. These distortions not only impair auditory experience but also severely hinder the performance of downstream speech processing systems. Traditional SE methods typically focus on addressing a single type of distortion—such as denoising, dereverberation, bandwidth extension, declipping, or packet loss concealment (PLC)—rendering them ill-suited for handling multiple co-occurring distortions in complex scenarios. To overcome this limitation, general speech enhancement (GSE) \cite{abs-2406-04660, LiuLKTZWHW22, abs-2406-02092, abs-2501-15417} has emerged as a research focus, aiming to develop unified frameworks capable of jointly restoring multiple types of distortions.

In recent years, deep learning-based SE approaches have achieved significant progress and can be broadly categorized into discriminative and generative paradigms. Discriminative methods, such as time-frequency masking and complex spectral mapping \cite{HaoSHL21, LuAL23}, typically excel in regression-oriented tasks like speech denoising and dereverberation by modeling deterministic mappings between noisy and clean speech. These techniques effectively suppress noise while preserving fine acoustic details. However, they are sensitive to the distribution of the training dataset, exhibit limited generalization in complex acoustic environments or unseen distortion types, and struggle with tasks requiring signal reconstruction. On the other hand, generative methods, including diffusion and autoregressive models, learn the latent distribution of clean speech, demonstrating superior adaptability in generation-oriented tasks such as bandwidth extension, declipping, and PLC. For instance, approaches leveraging neural audio codecs (NACs) and language models (LMs) can reconstruct high-quality speech using discrete speech tokens~\cite{WangZZLJZ024, yang24h, abs-2502-02942}. Nonetheless, these methods may introduce timbre distortion due to information loss and often suffer from high computational costs.

Although both discriminative and generative approaches offer distinct advantages, existing research is largely confined to a single paradigm, struggling to balance the precision required for regression tasks with the flexibility needed for generative tasks. Furthermore, there is limited exploration of cross-domain collaborative optimization between continuous signal processing and discrete token generation. Additionally, while NAC-based methods achieve high-quality reconstruction using discrete speech tokens, the inter-codebook dependencies within their hierarchical residual vector quantization (RVQ) structure remain under-utilized. This oversight results in cumulative quantization errors and acoustic inconsistencies, degrading overall performance.

To address these challenges, we propose OmniGSE, a unified general speech enhancement framework that synergistically integrates the strengths of both discriminative and generative approaches. Specifically, our method employs a generative SE strategy based on next-token prediction using NACs and LMs. It leverages the high-quality codebook priors of the pre-trained NAC and the powerful generative capabilities of autoregressive LMs. The framework operates in two distinct stages: In the first stage, a lightweight channel-split NAC-RoFormer network is introduced to perform discriminative enhancement on the pre-quantized continuous features extracted by the pre-trained NAC encoder. In the second stage, the enhanced high signal-to-noise ratio (SNR) pre-quantized features are used as conditioning inputs for autoregressive LMs to generate refined discrete speech tokens. By strategically combining the complementary strengths of both paradigms, OmniGSE achieves improved stability and performance through stage-wise handling of diverse distortion types.

Additionally, we designed a hierarchical LM architecture specifically for RVQ-based NACs, as illustrated in Figure~\ref{fig2}. This architecture comprises the following components:
\begin{itemize}
\item A RootLM that predicts universal features across all codebook layers.
\item Multiple BranchLMs, each responsible for predicting acoustic tokens for its corresponding layer based on outputs from both the RootLM and the preceding BranchLM.
\end{itemize}
This design is motivated by two key considerations:
\begin{itemize}
\item The RootLM learns shared high-level features (e.g., timbre, prosody) that serve as acoustic and semantic constraints. Meanwhile, the conditional dependency design of BranchLMs explicitly models progressive inter-layer acoustic relationships.
\item The use of separate BranchLMs mitigates inter-layer prediction conflicts (e.g., pattern contradictions between higher and lower codebooks). Additionally, the RootLM avoids redundant learning of low-level acoustic features, thereby improving parameter efficiency.
\end{itemize}

Compared to prior LM-based SE methods, our approach achieves high-fidelity and highly restorative enhancement without relying on additional pre-trained features (e.g., self-supervised learning (SSL) semantic features~\cite{yang24h,abs-2502-02942,kang2025}). This advantage stems from the high SNR conditional input provided by the first stage and the hierarchical LM architecture in the second stage.

In summary, our key contributions are summarized as follows:\\
\noindent \ding{113}~(1) We propose a two-stage GSE framework that integrates the complementary strengths of discriminative and generative approaches. By enabling cross-domain collaborative optimization of continuous signal features and discrete tokens, our framework achieves the precision of discriminative methods for tasks such as denoising and dereverberation, while also leveraging the flexibility of generative methods for tasks like bandwidth extension and declipping.\\
\noindent \ding{113}~(2) Exploiting the hierarchical nature of RVQ, we design a novel hierarchical LM architecture. The RootLM models universal acoustic features across codebook layers, providing high-level semantic constraints, while the BranchLM explicitly captures inter-layer progressive acoustic relationships. This design philosophy effectively reduces inter-layer prediction conflicts, ensures greater acoustic consistency, and enhances parameter efficiency.\\
\noindent \ding{113}~(3) Extensive evaluations on multiple GSE benchmarks demonstrate that OmniGSE surpasses existing models, achieving superior performance, particularly in complex scenarios involving compound distortions.

\section{Related Work}

\subsection{Language Model-Based Speech Enhancement Methods}

In recent years, LM-based SE methods have achieved significant progress, inspired by the successful application of large-scale LMs in cross-modal tasks. For instance, SELM~\cite{WangZZLJZ024} and LLaSE-G1~\cite{kang2025} employ k-means discrete tokens extracted by WavLM~\cite{ChenWCWLCLKYXWZ22} as intermediate representations and utilize LMs to perform autoregressive generation that maps noisy tokens to clean tokens. Similarly, MaskSR~\cite{abs-2406-02092} and AnyEnhance~\cite{abs-2501-15417} leverage masked generation techniques to enable the joint processing of multiple distortions, such as noise, reverberation, clipping, and bandwidth limitations. Additionally, GenSE~\cite{abs-2502-02942} adopts a two-stage framework: it first generates enhanced semantic tokens and subsequently reconstructs enhanced speech through a semantic-to-acoustic token generation process. These advancements highlight the growing potential of LM-based methods in addressing complex speech enhancement challenges.

\begin{figure*}[t]
    \centering
    \includegraphics[width=0.99\textwidth]{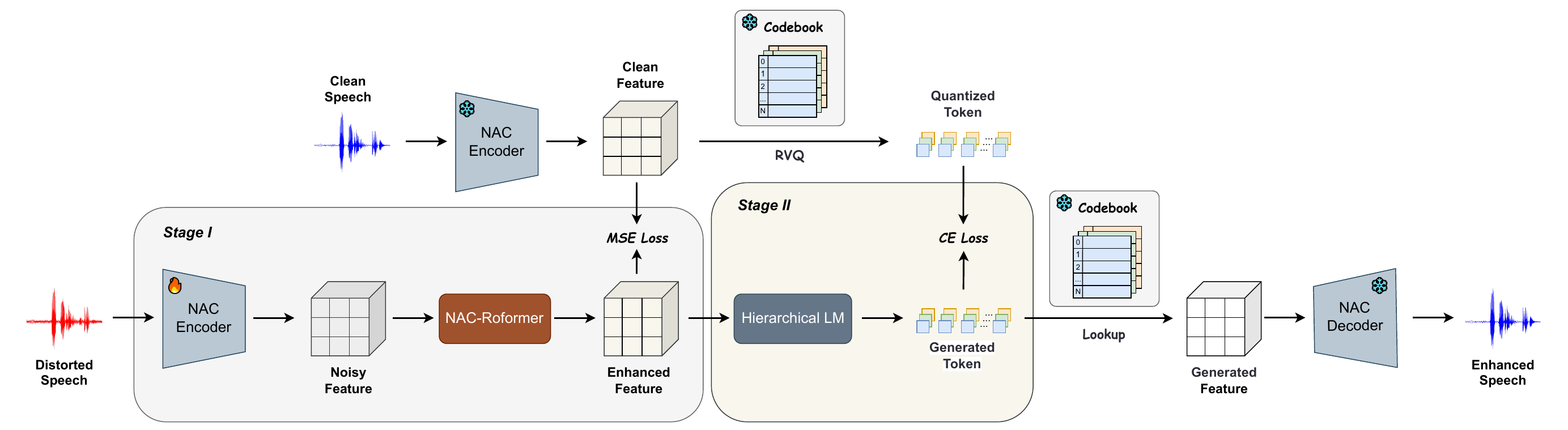}
    \caption{Workflow of the proposed OmniGSE framework.}
    \label{fig1}
\end{figure*}

\subsection{Neural Audio Codec-Based Speech Enhancement Methods}

Recent advancements in NAC technology have paved the way for innovative approaches to SE. Through large-scale pre-training, modern NAC models achieve high audio fidelity even at extreme compression rates, serving as a critical bridge between continuous speech signals and discrete language models. Within NACs, discrete codebooks—such as those implemented through vector quantization—typically encapsulate rich prior knowlege of clean speech characteristics. The vectors within these codebooks act as \textit{templates} or \textit{prototypes} of clean speech in the latent space, providing discrete priors that enhance robustness against various degradations while preserving fine-grained acoustic details for more natural enhancement outcomes. NAC-based SE methods reformulate SE as a token prediction task, broadly categorized into two paradigms: multi-codebook token prediction~\cite{abs-2406-02092, yang24h} and single-codebook token prediction~\cite{abs-2502-02942, kang2025}. Multi-codebook NACs, such as Encodec \cite{DefossezCSA23} and DAC \cite{KumarSLKK23}, exhibit superior audio reconstruction capabilities due to their hierarchical structure. In contrast, single-codebook NACs, including BigCodec~\cite{abs-2409-05377}, WavTokenizer~\cite{abs-2408-16532}, and X-codec2 \cite{abs-2502-04128}, offer lower prediction overhead, making them computationally efficient. \citeauthor{abs-2502-16240} \cite{abs-2502-16240} approached SE by enhancing pre-quantized features within NAC frameworks. Building upon this, our proposed OmniGSE further advances the framework by introducing a hierarchical LM architecture to convert the enhanced pre-quantized features into final acoustic tokens. This design not only retains the high performance of multi-codebook methods but also significantly reduces computational costs, striking a feasible balance between efficiency and quality.
\section{Method}
\subsection{Problem Formulation for General Speech Enhancement}
In this work, we focus on the following common speech distortions: noise, reverberation, clipping, bandwidth limitation, and packet loss. For a clean speech signal $x(t)$, each distorted speech signal $y(t)$ can be modeled as follows:
\begin{itemize}
\item \textbf{Noise}:
\begin{equation}
y(t)=x(t) + \alpha n(t),
\end{equation}
where $n(t)$ represents the noise interference, and $\alpha$ is a scaling factor determined by the SNR.
\item \textbf{Reverberation}:
\begin{equation}
y(t)=x(t) \ast h(t),
\end{equation}
where $h(t)$ denotes the room impulse response (RIR), and $\ast$ represents the convolution operation.
\item \textbf{Clipping}:
\begin{equation}
y(t)=\min (\max (x(t),a),b),
\end{equation}
where $a$ and $b$ are the clipping thresholds.
\item \textbf{Bandwidth Limitation}:
\begin{equation}
y(t)=\text{Upsample}_{f_\text{s}}(\text{Downsample}_{f_{\text{new}}}(x(t)))
\end{equation}
where $f_\text{s}$ is the original sampling rate, and $f_{\text{new}}$ is a randomly selected downsampling rate.
\item \textbf{Packet Loss}:
\begin{equation}
y(t)=x(t) \cdot \mathbb{I}(t \notin T_{\text{loss}}),
\end{equation}
where $T_{\text{loss}}$ represents the set of time intervals corresponding to packet loss.
\end{itemize}

To simulate real-world distorted speech, we apply these distortions sequentially according to the following rules:

\textbf{1. Noise} (added with 100\% probability) $\rightarrow$ \textbf{2. Reverberation} (applied with 50\% probability) $\rightarrow$ \textbf{3. Other distortions} (randomly selected with equal probability: clipping, bandwidth limitation, or packet loss).

\subsection{Overall Architecture}
The architecture of the proposed OmniGSE is illustrated in Figure~\ref{fig1}. OmniGSE operates in two stages. In the first stage, a channel-split NAC-RoFormer is employed to pre-enhance high-dimensional features encoded by the NAC encoder. The NAC encoder is fine-tuned to adapt to distorted speech inputs. In the second stage, the enhanced pre-quantized features are fed into our proposed hierarchical LM as conditioning inputs. The model autoregressively generates tokens for each layer of the RVQ codebook. Finally, the enhanced speech is reconstructed using the NAC decoder. Detailed descriptions of the first and second stages will be illustrated in Secs.~\ref{3.3} and \ref{3.4}, respectively.

\subsection{Stage \uppercase\expandafter{\romannumeral1}: Continuous Feature Enhancement}
 \label{3.3}

To capture rich speech information while improving reconstruction quality, NACs typically utilize high-dimensional encoding features (e.g., 1024-dimensional in DAC \cite{KumarSLKK23}), which leads to high computational complexity. To reduce computational costs while addressing the over-smoothing issue associated with global attention \cite{ZaheerGDAAOPRWY20, XiaPSLH22}, we propose a dual-path channel-split NAC-RoFormer tailored to the characteristics of NACs. This architecture groups channels for dimensionality reduction and collaboratively computes local and global attention in a dual-path manner.

Specifically, given the NAC-encoded features $F_{\text{enc}}\in\mathbb{R}^{D\times T}$, where $D$ represents the feature dimension and $T$ denotes the number of time steps, we first uniformly split $F_{\text{enc}}$ along the feature dimension $D$ into $G$ non-overlapping groups. This results in grouped features $F^{\prime}\in \mathbb{R}^{G \times D_{\text{group}} \times T}$, where $D_{\text{group}}$ is the feature dimension per group ($D = G \times D_{\text{group}}$). Then, for $F^{\prime}$, we use the dual-path method \cite{YuL23,LuWKH24} to compute self-attention on the temporal axis $T$ within each group and on the channel group axis $G$ across groups using the RoFormer \cite{SuALPBL24}. Finally, the grouped features are merged along the channel dimension and concatenated back to their original dimension to produce the enhanced continuous features $F_{\text{enh}}$. We employ the Snake activation function \cite{LiuHU20} to introduce periodic inductive bias, while adapting the activation layer of the DAC encoder to maintain consistent output amplitude \cite{yip2024towards}.

\textbf{Training.} To train our continuous NAC feature enhancement network, we introduce a teacher NAC network $N_{\text{tea}}$ that provides the NAC-encoded embeddings of clean speech as learning targets $F_{\text{tea}}$. We employ the mean squared error (MSE) as the loss function:
\begin{equation}
\mathcal{L}_{\text{emb}}=\text{MSE}(F_{\text{enh}}, F_{\text{tea}}).
\end{equation}

Since the NAC has been pre-trained exclusively on large-scale clean speech data, it may exhibit pattern mismatch issues when encoding distorted speech, resulting in unstable performance. To address this, we simultaneously fine-tune the NAC encoder during the training of our continuous NAC feature enhancement network, enabling it to better adapt to various types of distorted speech inputs.

\begin{figure}[t]
    \centering
    \includegraphics[width=0.49\textwidth]{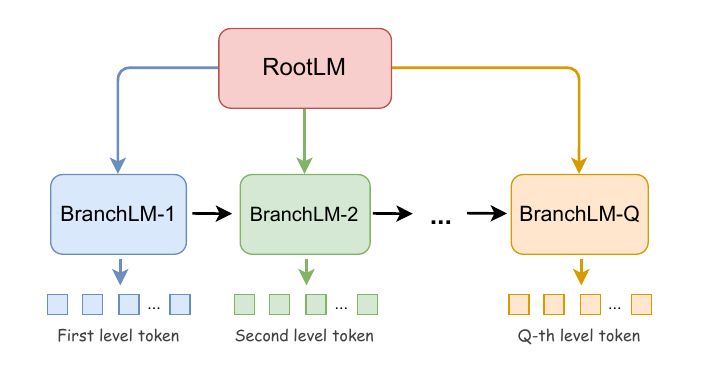}
    \caption{Topology of the hierarchical language model.}
    \label{fig2}
\end{figure}

\subsection{Stage \uppercase\expandafter{\romannumeral2}: Discrete Token Generation}
 \label{3.4}

Following the continuous feature enhancement in the first stage, we obtain high-SNR{\footnote{Here, the SNR is defined in the latent space rather than the signal space.} pre-quantized features $F_{\text{enh}}$. While $F_{\text{enh}}$ demonstrates effective pre-enhancement for distortions like additive noise \cite{abs-2502-16240}, it still struggles to reconstruct missing content relative to the target clean audio-such as certain frequency bands, amplitude values, or temporal frames. To address this limitation, we introduce the second stage: NAC discrete token generation. This stage leverages the powerful multi-modal generation capability of autoregressive LMs to handle such content-missing distortions more effectively.

The RVQ structure in NACs progressively refines speech representations through multi-level codebooks. Lower-level codebooks capture global acoustic features (e.g., speaker identity, fundamental frequency contour), while higher-level codebooks supplement fine-grained details (e.g., high-frequency harmonics, transient components). To explicitly model this hierarchical dependency, we designed a hierarchical LM, comprising a \textit{RootLM} shared across all codebook levels and multiple \textit{BranchLMs} that are independent between codebooks (the number of BranchLMs matches the number of RVQ codebook levels), as illustrated in Figure \ref{fig2}. The workflow proceeds as follows:

\begin{itemize}
\item The RootLM takes the enhanced continuous feature $F_{\text{enh}}$ from the first stage as the conditional input. It autoregressively generates $H_{\text{root}}$, which encapsulates universal acoustic features shared across all codebooks (such as timbre and prosody) and provides semantic constraints. This serves as acoustic and semantic guidance for token prediction at subsequent levels.
\item For each level $l$, the corresponding BranchLM takes $H_{\text{root}}$ and the NAC token sequence $z_{l-1}$ from the previous level $l-1$ as conditions to predict the discrete NAC tokens $z_l$ at the current level. This approach explicitly captures the progressive acoustic relationship across levels. Notably, each BranchLM is independently parameterized to avoid interference between levels.
\end{itemize}

\textbf{Training.} To train the hierarchical LM, we utilize the codebook tokens of clean speech provided by the teacher NAC network $N_{\text{tea}}$ as learning targets. Additionally, we adopt a teacher-forcing training strategy for predicting tokens at different codebook levels to mitigate error accumulation and enhance training stability. Specifically, during training, the ground-truth token sequence of the $(l-1)$-th layer, provided by the teacher NAC network $N_{\text{tea}}$, is used as conditional input for the $l$-th layer BranchLM. During inference, the predicted token sequence from the $(l-1)$-th layer BranchLM serves as the conditional input for the $l$-th layer BranchLM. We employ cross-entropy (CE) loss to jointly optimize the RootLM and all BranchLMs:
\begin{equation}
\mathcal{L}_{\text{code}}= -\sum_{l=1}^Q \sum_{t=1}^T \log p(z^l_t \mid z^l_{<t}, H_{\text{root}}, \tilde{z}^{l-1}),
\end{equation}
where $Q$ represents the number of codebooks, $z^l_t$ denotes the predicted token at each time step $t$ for the $l$-th codebook, and $\tilde{z}$ corresponds to the ground-truth tokens provided by the teacher network $N_{\text{tea}}$.

\section{Experiments}

\begin{table*}[th]
\caption{Results on the Interspeech 2020 DNS Challenge blind test set for denoising and dereverberation. ``D'' represents discriminative methods, while ``G'' represents generative methods. Bold and underlined numbers indicate the best and second-best results, respectively.}
\centering
\small
\begin{tabular}{llrrrrrr|rrrrrr}
\hline
\multicolumn{1}{c}{\multirow{2}{*}{Method}} & \multirow{2}{*}{Type} & \multicolumn{6}{c|}{No Reverb}                                                                                                                                                                                    & \multicolumn{6}{c}{With Reverb}                                                                                                                                                                                                \\ \cline{3-14} 
\multicolumn{1}{c}{}                        &                       & \multicolumn{1}{l}{SIG $\uparrow$} & \multicolumn{1}{l}{BAK $\uparrow$} & \multicolumn{1}{l}{OVRL $\uparrow$} & \multicolumn{1}{l}{NISQA $\uparrow$} & \multicolumn{1}{l}{SBS $\uparrow$} & \multicolumn{1}{l|}{SIM $\uparrow$} & \multicolumn{1}{l}{SIG $\uparrow$} & \multicolumn{1}{l}{BAK $\uparrow$} & \multicolumn{1}{l}{OVRL $\uparrow$} & \multicolumn{1}{l}{NISQA $\uparrow$} & \multicolumn{1}{l}{SBS $\uparrow$} & \multicolumn{1}{l}{SIM $\uparrow$} \\ \hline
DEMUCS                                      & D                     & $3.533$                            & $4.157$                            & $3.310$                             & $3.742$                              & $0.877$                            & $0.984$                            & $2.937$                            & $3.844$                            & $2.615$                             & $2.188$                              & $0.725$                            & $0.930$                            \\
FRCRN                                       & D                     & $3.574$                            & $4.154$                            & $3.332$                             & $4.495$                              & $\mathbf{0.914}$                   & $\mathbf{0.993}$                   & $2.933$                            & $2.923$                            & $2.279$                             & $2.270$                              & $0.783$                            & {\ul $0.966$}                   \\
VoiceFixer                                  & D                     & $3.500$                            & $4.110$                            & $3.250$                             & $4.270$                              & $-$                                & $0.960$                            & $3.430$                            & $4.020$                            & $3.130$                             & $\mathbf{3.820}$                         & $-$                                & $0.910$                            \\
TF-GridNet                                   & D                     & $3.539$                            & $4.047$                            & $3.268$                             & $4.347$                              & $0.902$                            & $0.675$                            & $3.110$                            & $3.225$                            & $2.510$                             & $2.614$                              & $\mathbf{0.840}$                   & $0.686$                            \\ \hline
SELM                                        & G                     & $3.508$                            & $4.096$                            & $3.258$                             & $-$                                  & $-$                                & $-$                                & $3.160$                            & $3.577$                            & $2.695$                             & $-$                                  & $-$                                & $-$                                \\
MaskSR                                      & G                     & $3.616$                            & {\ul $4.183$}                      & $3.393$                             & $4.754$                              & $0.875$                            & $0.983$                            & $3.396$                            & $4.043$                            & $3.085$                             & $3.353$                              & $0.701$                            & $0.946$                            \\
GenSE                                       & G                     & $3.650$                            & $4.180$                            & {\ul $3.430$}                       & $-$                                  & $-$                                & $-$                                & $3.490$                            & $3.730$                            & $3.190$                             & $-$                                  & $-$                                & $-$                                \\
AnyEnhance                                  & G                     & $3.640$                            & $4.179$                            & $3.418$                             & {\ul $4.821$}                        & $0.907$                            & $0.988$                      & $3.500$                            & $4.040$                            & $3.204$                             & $3.722$                              & $0.738$                            & $0.951$                            \\
LLaSE-G1                                    & G                     & {\ul $3.660$}                      & $4.170$                            & $3.420$                             & $-$                                  & $-$                                & $-$                                & {\ul $3.590$}                            & {\ul $4.100$}                            & $\mathbf{3.330}$                       & $-$                                  & $-$                                & $-$                                \\ \hline
$\text{OmniGSE}_{\text{wb}}$                                 & D+G                   & $\mathbf{3.706}$                   & $\mathbf{4.250}$                   & $\mathbf{3.444}$                    & $\mathbf{4.828}$                     & {\ul $0.910$}                      & {\ul $0.990$}                            & $\mathbf{3.627}$                      & $\mathbf{4.167}$                   & {\ul $3.314$}                             & {\ul $3.809$}                              & {\ul $0.803$}                      &  $\mathbf{0.980}$                      \\ \hline
\end{tabular}
\label{tab1}
\end{table*}

\begin{figure*}[th]
    \centering
    \includegraphics[width=0.99\textwidth]{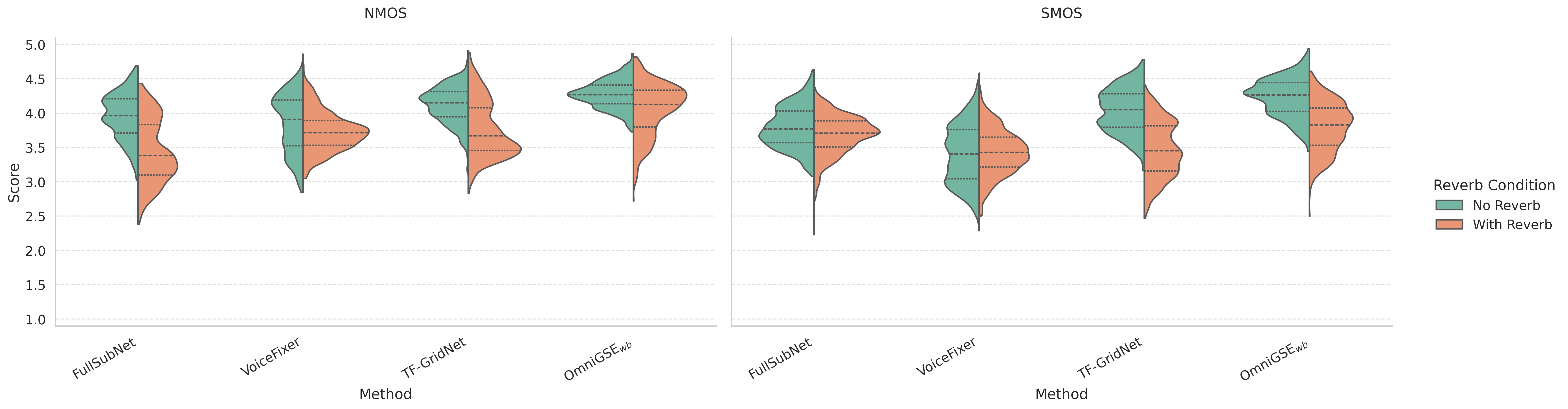}
    \caption{Violin plots of NMOS and SMOS scores for various methods on the Interspeech 2020 Challenge blind test set. The black dashed lines within the violin plots represent the quartile boundaries.}
    \label{fig3}
\end{figure*}

\begin{table*}[th]
\caption{Results for speech super-resolution on the Voicefixer SR test set.}
\centering
\small
\begin{tabular}{llrrrrrr}
\hline
\multicolumn{1}{c}{\multirow{2}{*}{Method}} & \multirow{2}{*}{Type} & \multicolumn{6}{c}{SR}                                                                                          \\ \cline{3-8} 
\multicolumn{1}{c}{}                        &                       & SIG $\uparrow$   & BAK $\uparrow$   & OVRL $\uparrow$  & NISQA $\uparrow$ & SBS $\uparrow$   & SIM $\uparrow$  \\ \hline
VoiceFixer                                  & D                     & $3.405$          & $4.029$          & $3.110$          & $4.131$          & $0.873$          & $0.882$          \\ \hline
AudioSR                                     & G                     & {\ul $3.492$}    & $4.002$          & {\ul $3.180$} & $4.255$          & $0.913$          & $0.911$          \\
MaskSR                                      & G                     & $3.464$          & $4.028$          & $3.154$          & {\ul $4.352$}    & $0.925$          & {\ul $0.939$}    \\
AnyEnhance                                  & G                     & $3.449$          & {\ul $4.063$}    & $3.156$          & $4.201$          & $\mathbf{0.941}$ & $\mathbf{0.943}$ \\ \hline
$\text{OmniGSE}_{\text{fb}}$                                 & D+G                   & $\mathbf{3.498}$ & $\mathbf{4.137}$ & $\mathbf{3.181}$    & $\mathbf{4.365}$ & {\ul $0.930$}    & $0.935$          \\ \hline
\end{tabular}
\label{tab2}
\end{table*}

\begin{table*}[th]
\caption{Results for general speech restoration on the Voicefixer GSR test set.}
\centering
\small
\begin{tabular}{llrrrrrr}
\hline
\multicolumn{1}{c}{\multirow{2}{*}{Method}} & \multirow{2}{*}{Type} & \multicolumn{6}{c}{GSR}                                                                                         \\ \cline{3-8} 
\multicolumn{1}{c}{}                        &                       & SIG $\uparrow$   & BAK $\uparrow$   & OVRL $\uparrow$  & NISQA $\uparrow$ & SBS $\uparrow$   & SIM $\uparrow$   \\ \hline
NSNet2                                      & D                     & $3.011$          & $3.969$          & $2.785$          & $3.433$          & $0.728$          & $0.615$          \\
VoiceFixer                                  & D                     & $3.299$          & $3.971$          & $3.003$          & $4.160$          & $0.797$          & $0.882$          \\
TF-GridNet                                  & D                     & $3.253$          & $3.906$          & $2.945$          & $3.643$          & $0.782$          & $0.613$          \\ \hline
MaskSR                                      & G                     & {\ul $3.408$} & $4.041$          & $3.122$          & $\mathbf{4.335}$ & {\ul $0.832$}    & $0.916$          \\
AnyEnhance                                  & G                     & $3.406$    & {\ul $4.073$}    & {\ul $3.136$}    & {\ul $4.308$}    & $0.829$          & {\ul $0.924$}    \\ \hline
$\text{OmniGSE}_{\text{fb}}$                                     & D+G                   & $\mathbf{3.420}$          & $\mathbf{4.108}$ & $\mathbf{3.149}$ & $4.293$          & $\mathbf{0.912}$ & $\mathbf{0.938}$ \\ \hline
\end{tabular}
\label{tab3}
\end{table*}

\begin{table}[th]
\caption{Results of packet loss concealment on the Interspeech 2022 PLC blind test set.}
\centering
\small
\begin{tabular}{llrr}
\hline
\multicolumn{1}{c}{\multirow{2}{*}{Method}} & \multirow{2}{*}{Type} & \multicolumn{2}{c}{PLC}             \\ \cline{3-4} 
\multicolumn{1}{c}{}                        &                       & OVRL $\uparrow$ & PMOS $\uparrow$ \\ \hline
KuaishouNet                                 & D                     & $-$             & $4.27$            \\
LPCNet                                      & D                     & $3.09$          & $3.74$            \\
PLCNet                                      & D                     & $-$             & $3.83$            \\
BS-PLCNet                                   & D                     & {\ul $3.20$}    & {\ul $4.29$}      \\ \hline
LLaSE-G1                                    & G                     & $3.03$          & $3.68$            \\ \hline
$\text{OmniGSE}_{\text{fb}}$                                      & D+G                   & $\mathbf{3.25}$ & $\mathbf{4.33}$   \\ \hline
\end{tabular}
\label{tab4}
\end{table}

\begin{figure*}[th]
    \centering
    \includegraphics[width=0.99\textwidth]{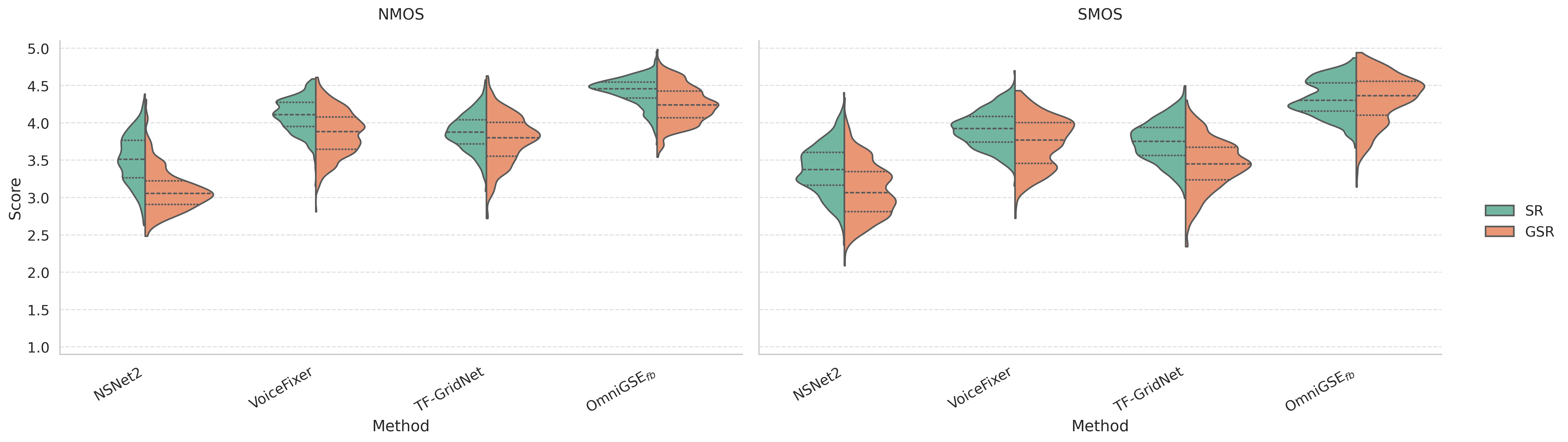}
    \caption{Violin plots of NMOS and SMOS scores for various methods on the Voicefixer SR and Voicefixer GSR test sets.}
    \label{fig4}
\end{figure*}

\begin{table*}[th]
\caption{Ablation study results on the Interspeech 2020 Challenge blind test set and the Voicefixer GSR test set.}
\centering
\small
\begin{tabular}{llrrrrrr}
\hline
\multicolumn{1}{c}{\multirow{2}{*}{Exp.}} & \multicolumn{1}{c}{\multirow{2}{*}{Method}} & \multicolumn{2}{c}{No Reverb}       & \multicolumn{2}{c}{With Reverb}     & \multicolumn{2}{c}{GSR}             \\ \cline{3-8} 
\multicolumn{1}{c}{}                      & \multicolumn{1}{c}{}                        & OVRL $\uparrow$  & NISQA $\uparrow$ & OVRL $\uparrow$  & NISQA $\uparrow$ & OVRL $\uparrow$  & NISQA $\uparrow$ \\ \hline
(a)                                       & Baseline                                    & $\mathbf{3.444}$ & $\mathbf{4.828}$ & $\mathbf{3.314}$ & $\mathbf{3.809}$ & $\mathbf{3.149}$ & $\mathbf{4.293}$ \\
(b)                                       & w/o Stage \uppercase\expandafter{\romannumeral1}                                 & $3.112$          & $4.412$          & $2.998$          & $3.502$          & $2.921$          & $4.065$          \\
(c)                                       & w/o Stage \uppercase\expandafter{\romannumeral2}                                 & $3.201$          & $4.595$          & $3.088$          & $3.581$          & $2.887$          & $3.912$          \\
(d)                                       & w/o NAC-Roformer                            & $3.085$          & $4.423$          & $2.972$          & $3.462$          & $2.801$          & $3.786$          \\
(e)                                       & w/o Hierarchical LM                         & $3.056$          & $4.325$          & $2.945$          & $3.402$          & $2.765$          & $3.698$          \\
(f)                                       & w/o FT Encoder                              & $3.092$          & $4.318$          & $3.001$          & $3.435$          & $2.712$          & $3.721$          \\
(g)                                       & w/o Teacher-forcing                         & $3.228$          & $4.632$          & $3.102$          & $3.620$          & $2.973$          & $3.945$          \\ \hline
\end{tabular}
\label{tab5}
\end{table*}

\begin{figure}[th]
    \centering
    \includegraphics[width=0.49\textwidth]{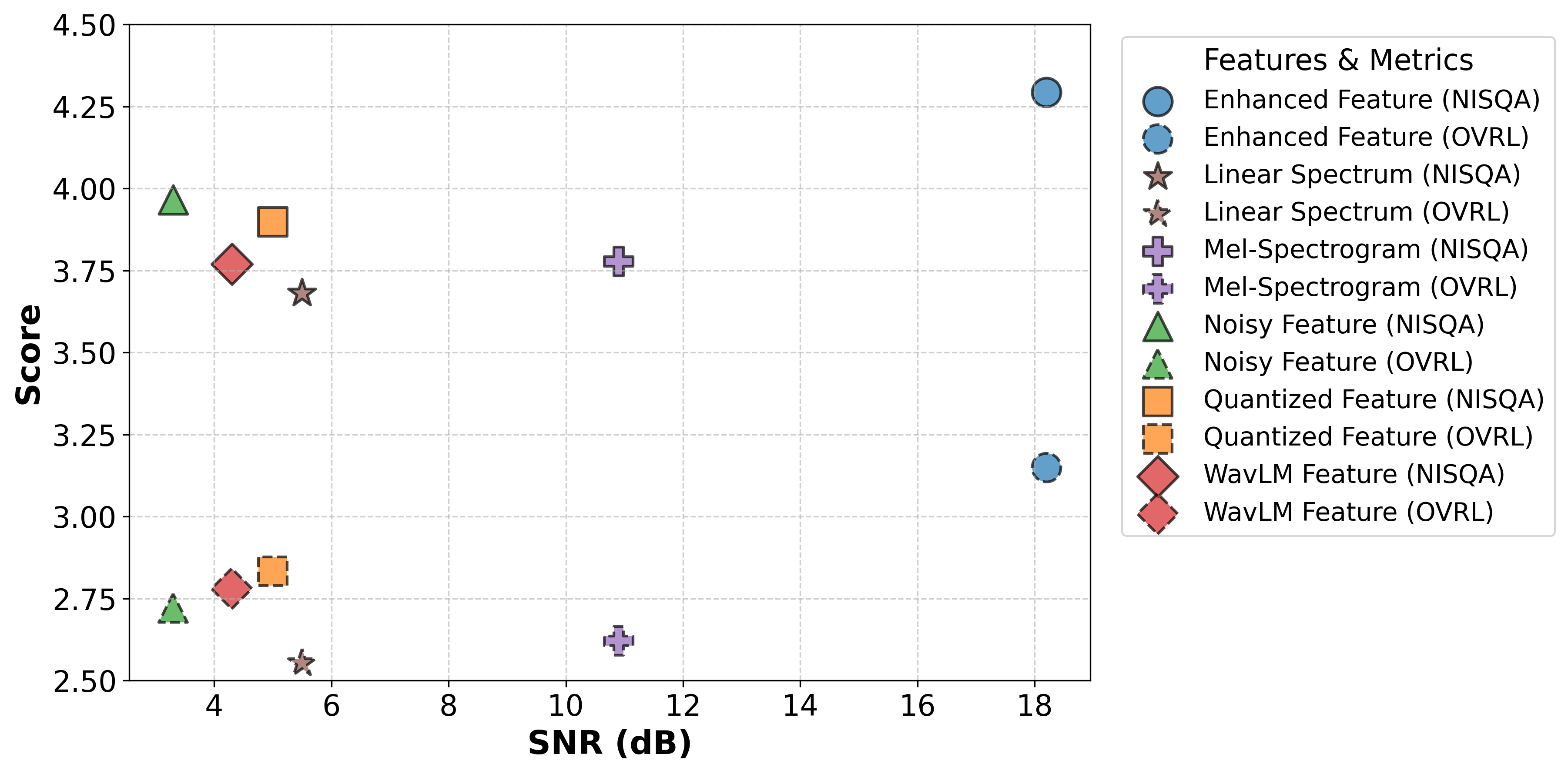}
    \caption{Results and SNR of different input features on the Voicefixer GSR test set.}
    \label{fig5}
\end{figure}

\subsection{Datasets}

To thoroughly evaluate our method, we conducted experiments on both wideband speech (16 kHz sampling rate) and full-band speech (44.1 kHz sampling rate). For wideband speech, we utilized clean speech data from the DNS5 Challenge dataset \cite{dubey2024icassp}, LibriTTS \cite{ZenDCZWJCW19}, and VCTK \cite{veaux2017cstr}, all resampled to 16 kHz. For full-band speech, we employed 44.1 kHz and 48 kHz clean speech data from the DNS5 Challenge dataset, HiFi-TTS \cite{BakhturinaLGZ21}, and VCTK \cite{veaux2017cstr}, with all samples resampled to 44.1 kHz. Additionally, we fine-tuned our model on a private high-fidelity speech dataset. The full-band noise data was sourced from the DNS Challenge, FSD50K \cite{FonsecaFPFS22}, and WHAM! \cite{WichernAFZMCMR19}. Noise mixing was performed with the SNR ranging from -5 dB to 20 dB. Room impulse response (RIR) data provided by the DNS Challenge were used to simulate reverberation. All noise and RIR samples were resampled to 16 kHz (for wideband models) or 44.1 kHz (for full-band models). Each speech segment was truncated to a duration of 2 seconds. To simulate bandwidth limitation, we randomly downsampled the 16 kHz speech samples to 2 kHz, 4 kHz, and 8 kHz, and the 44.1 kHz speech samples to 2 kHz, 4 kHz, 8 kHz, 16 kHz, and 24 kHz. All training data were generated on the fly during training.

\subsection{Model Configuration}

Since the NAC is typically trained on speech at a specific sampling rate, we trained separate models for wideband speech (16 kHz) and full-band speech (44.1 kHz), referred to as ($\text{OmniGSE}_{\text{wb}}$) and ($\text{OmniGSE}_{\text{fb}}$), respectively. For both models, we utilized the pre-trained DAC \cite{KumarSLKK23} as the NAC, configured for 16 kHz and 44.1 kHz sampling rates. The wideband model employs 12 codebooks ($Q=12$), while the full-band model uses 9 codebooks ($Q=9$). Each codebook has a size of $N=1024$, and the encoded feature dimension is $D=1024$. The channel-split NAC-RoFormer consists of 12 layers of RoFormer, alternately processing along the temporal and channel axes. We divided the channels into $G=64$ groups, with each group containing $D_{\text{group}}=16$ channel features. Both the temporal and channel RoFormer modules have a feature dimension of 64 and utilize 8 attention heads. The total number of parameters in the model is approximately 7.6M. 

For the hierarchical LM, we adopted a LLaMA-style Transformer \cite{abs-2302-13971} as the backbone architecture. Specifically, the RootLM and BranchLMs (excluding the first level) consist of a 6-layers and a 1-layer Transformer, respectively. Given that the first-level codebook captures the majority of the speech information, we enhanced its modeling capacity by employing a 2-layer Transformer for the first-level BranchLM. Each Transformer layer features a hidden layer dimension of 1024, an intermediate size of 4096, 16 attention heads, and a dropout rate of 0.1. The wideband and full-band models contain approximately 1.23B and 0.97B parameters, respectively. All models were trained using the AdamW optimizer \cite{LoshchilovH19} with betas set to (0.8, 0.999) and an initial learning rate of $5 \times 10^{-4}$, followed by exponential decay. The training process involved separate optimization of the first and second stages in a progressive manner.

\subsection{Evaluation Metrics}

To comprehensively evaluate the performance of OmniGSE, we employed a variety of objective and subjective metrics to assess the perceptual quality, content restoration, and speaker characteristic retention of the enhanced speech. Specifically, the objective metrics include:
\begin{itemize}
\item \textbf{DNSMOS} \cite{ReddyGC22}: A no-reference perceptual quality estimator that outputs three scores ranging from 1 to 5, including speech quality (SIG), background noise quality (BAK), and overall audio quality (OVRL).
\item \textbf{NISQA} \cite{MittagNC021}: A no-reference perceptual quality estimator that outputs a single score ranging from 1 to 5, representing the overall quality of the speech signal.
\item \textbf{PLCMOS (PMOS)} \cite{DienerPSSAC23}: A metric designed to evaluate the quality of speech enhanced by PLC algorithms, outputting a single score ranging from 1 to 5.
\item \textbf{SpeechBERTScore (SBS)} \cite{abs-2401-16812}: A metric used to measure semantic similarity between the enhanced speech and the reference speech. We utilized the pre-trained HuBERT-base model\footnote{\url{https://huggingface.co/facebook/hubert-base-ls960}} \cite{HsuBTLSM21} to extract semantic features.
\item \textbf{Speaker Similarity (SIM)}: To evaluate speaker characteristic retention, we extracted speaker embeddings using the pre-trained WavLM model\footnote{\url{https://huggingface.co/microsoft/wavlm-base-plus-sv}} \cite{ChenWCWLCLKYXWZ22} and calculated the cosine similarity between the enhanced and reference signals.
\end{itemize}
The subjective metrics include:
\begin{itemize}
\item \textbf{Naturalness Mean Opinion Score (NMOS)} and \textbf{Similarity Mean Opinion Score (SMOS)}: Participants were asked to evaluate the naturalness and speaker similarity of the enhanced speech on a scale from 1 to 5.
\end{itemize}

\subsection{Comparison with State-of-the-Art Methods}

In this section, we conducted a detailed comparison of our proposed OmniGSE with other leading baseline methods for the general speech enhancement task. Sec.~\ref{4.4.1} focuses on evaluating performance for traditional types of speech distortions, such as noise and reverberation. Sec.~\ref{4.4.2} extends the comparison to include distortions involving loss of speech information, such as bandwidth limitation, clipping, and packet loss, in addition to the aforementioned distortion types.

\subsubsection{Results on speech denoising and dereverberation}
\label{4.4.1}

We first conducted an objective evaluation of our method for denoising and dereverberation tasks in traditional speech enhancement. The blind test set from the Interspeech 2020 Challenge \cite{ReddyGCBCDMAABR20} was used as the benchmark, which consists of two subsets: one without reverberation (No Reverb) and one with reverberation (With Reverb). Given that the speech sampling rate in this test set is 16 kHz, we evaluated the performance using the wideband model $\text{OmniGSE}_{\text{wb}}$. The baseline models included discriminative methods such as DEMUCS \cite{DefossezSA20}, FRCRN \cite{ZhaoMWG22}, VoiceFixer \cite{LiuLKTZWHW22}, and TF-GridNet \cite{WangCCLKW23}, as well as generative methods such as SELM \cite{WangZZLJZ024}, MaskSR \cite{abs-2406-02092}, GenSE \cite{abs-2502-02942}, AnyEnhance \cite{abs-2501-15417}, and LLaSE-G1 \cite{kang2025}. The results, presented in Table \ref{tab1}, demonstrate that our proposed $\text{OmniGSE}_{\text{wb}}$ achieves competitive performance in both denoising and dereverberation tasks.

For the subjective evaluation of denoising and dereverberation tasks, we collected NMOS and SMOS scores from participants for various methods on the two blind test sets of the Interspeech 2020 Challenge. The results were visualized using violin plots. We selected open-source models that performed well on these datasets—namely FullSubNet \cite{HaoSHL21}, VoiceFixer \cite{LiuLKTZWHW22}, and TF-GridNet \cite{WangCCLKW23}—as baseline methods. As shown in Figure \ref{fig3}, our $\text{OmniGSE}_{\text{wb}}$ significantly outperformed the baseline methods in terms of both NMOS and SMOS scores. These results indicate that the speech enhanced by our method achieves superior naturalness and effectively retains speaker characteristics.

\subsubsection{Results on speech restoration} 
\label{4.4.2}

To objectively evaluate the restoration performance of our method when dealing with multiple types of distortions, we conducted experiments on the Voicefixer SR and Voicefixer GSR full-band test sets \cite{LiuLKTZWHW22} for speech super-resolution and general speech restoration tasks, respectively. The Voicefixer SR test set contains speech distortions limited to bandwidth restrictions, while the Voicefixer GSR test set includes a broader range of distortions, such as noise, reverberation, clipping, and bandwidth limitation. Given that the target speech sampling rate in both test sets is 44.1 kHz, we evaluated the performance using the full-band model $\text{OmniGSE}_{\text{fb}}$. The baseline methods included discriminative approaches such as NSNet2 \cite{BraunT20}, VoiceFixer \cite{LiuLKTZWHW22}, and TF-GridNet \cite{WangCCLKW23}, as well as generative approaches like AudioSR \cite{Liu0T0P24}, MaskSR \cite{abs-2406-02092}, and AnyEnhance \cite{abs-2501-15417}. The results for the SR and GSR tasks are presented in Tables \ref{tab2} and \ref{tab3}, respectively. These results indicate that our proposed $\text{OmniGSE}_{\text{fb}}$ outperformed most baseline methods across key metrics, highlighting the effectiveness of our two-stage approach. Notably, our method not only removes interfering components from distorted speech but also successfully restores missing content, showcasing its robustness and versatility in speech restoration tasks.

Additionally, to evaluate the capability of our method in handling packet loss distortions, we conducted assessments on the Interspeech 2022 PLC blind test set \cite{DienerSBSAC22}. The speech sampling rate in this test set is 48 kHz. We first downsampled the data to 44.1 kHz and then evaluated it using the full-band model $\text{OmniGSE}_{\text{fb}}$. The baseline methods included discriminative approaches such as KuaishouNet \cite{LiZZGY22}, LPCNet \cite{ValinMMTKSK22}, PLCNet \cite{LiuSYYW22}, and BS-PLCNet \cite{ZhangSXHXX24}, as well as the generative method LLaSE-G1 \cite{kang2025}. Notably, except for our method and LLaSE-G1, all other models require prior indication of which frames have packet loss through lossy labels. The results for the PLC task are presented in Table \ref{tab4}. Remarkably, our method surpasses previous informed PLC approaches even in the more challenging blind PLC scenario.

For the subjective evaluation of the speech restoration task, we collected the NMOS and SMOS scores from participants for various methods on the Voicefixer SR and Voicefixer GSR test sets. The results were visualized using violin plots. The baseline methods included NSNet2 \cite{BraunT20}, VoiceFixer \cite{LiuLKTZWHW22}, and TF-GridNet \cite{WangCCLKW23}. As shown in Figure \ref{fig4}, our $\text{OmniGSE}_{\text{fb}}$ significantly outperformed the baseline methods in both NMOS and SMOS scores. These results indicate that the distorted speech restored by our method achieves high naturalness while effectively preserving speaker similarity.

\subsection{Ablation Study}

To evaluate the effectiveness of the improvements proposed in our method, we conducted ablation studies in this section. We performed comparisons on two test sets from the Interspeech 2020 Challenge, which include speech without reverberation (No Reverb) and with reverberation (With Reverb), as well as on the full-band Voicefixer GSR test set. The baseline models used were $\text{OmniGSE}_{\text{wb}}$ and $\text{OmniGSE}_{\text{fb}}$, respectively. The results are presented in Table \ref{tab5}.

\subsubsection{Ablation study on the two-stage approach}

To assess the contribution of our proposed two-stage approach, we conducted experiments by selectively removing either the continuous feature enhancement process in the first stage or the discrete token prediction process in the second stage. Specifically, when the first stage was omitted, the output features from the pre-trained NAC encoder were directly fed into the hierarchical LM in the second stage. Conversely, when the second stage was omitted, the enhanced features produced by the NAC-RoFormer were quantized and decoded by the pre-trained DAC quantizer and decoder to generate the enhanced speech. The corresponding results are shown in Exp. (b) and (c) of Table \ref{tab5}. The findings reveal that removing the continuous feature enhancement process in the first stage leads to a significant decline in the model's performance for denoising and dereverberation tasks. Similarly, omitting the discrete token prediction process in the second stage results in a marked reduction in the model's ability to restore missing speech content. These results confirm that the first stage primarily addresses the regression-oriented task, such as denoising and dereverberation, at the continuous feature level, while the second stage focuses on generation-oriented tasks, such as speech restoration, at the discrete token level. Our baseline method effectively handles both types of distortions through cross-domain collaborative enhancement across the two stages.

\subsubsection{Ablation study on model architecture}

First, to verify the effectiveness of the channel-split NAC-RoFormer, we replaced it with a standard Transformer, building upon Exp. (c) in Table \ref{tab5}, similar to the approaches described in \cite{yip2024towards,abs-2502-16240}. The results are presented in Exp. (d) of Table \ref{tab5}. These findings indicate that our channel-split NAC-RoFormer benefits from dual-path modeling along the channel and temporal axes, as well as the use of Rotated Position Embedding (RoPE) \cite{SuALPBL24}. This design not only significantly improves enhancement performance but also reduces computational cost.

Second, to validate the effectiveness of our proposed hierarchical LM, we compared it with an alternative approach that uses the same LM to predict tokens across all levels, akin to the method in \cite{chen2025neural}. Specifically, an autoregressive LM predicts tokens for the first-level codebook, while another non-autoregressive LM predicts tokens for all remaining levels. The results are shown in Exp. (e) of Table \ref{tab5}. The results reveal that without our hierarchical LM, all metrics decline significantly, even falling below the performance in Exp. (c), where the second stage was entirely omitted. This indicates that using a single LM to predict tokens at different levels introduces severe hierarchical pattern conflicts, leading to a marked degradation in overall performance.

\subsubsection{Ablation study on training methods}

To verify the effectiveness of fine-tuning the DAC encoder during the first stage of training, we fixed the pre-trained DAC encoder based on Exp. (c) in Table \ref{tab5}. The results, shown in Exp. (f) of Table \ref{tab5}, indicate that the DAC encoder, which is pre-trained on clean speech, exhibits a significant decline in encoding performance without fine-tuning on distorted speech inputs.

Additionally, to assess the effectiveness of teacher-forcing learning for training multi-level BranchLMs, we replaced the ground-truth token sequence from the $(l-1)$-th layer with the predicted token sequence as the conditioning input for the $l$-th layer BranchLM. The results are presented in Exp. (g) of Table \ref{tab5}. These findings demonstrate that teacher-forcing learning effectively mitigates error accumulation when predicting multi-layer codebook tokens, thereby enhancing the prediction accuracy of the BranchLM for the RVQ NAC codes.

\subsubsection{Comparison of different conditional features}

To verify the effectiveness of using the pre-quantized features enhanced in the first stage as the conditional input for the second stage, we conducted experiments with various alternative conditional inputs in place of the pre-enhanced features (Enhanced Feature) from the first stage. These alternatives included:
\begin{itemize}
\item Features obtained by quantizing the enhanced features from the first stage using the DAC quantizer (Quantized Feature).
\item Noisy features without enhancement by the NAC-RoFormer (Noisy Feature).
\item Semantic features encoded by WavLM \cite{ChenWCWLCLKYXWZ22} from the distorted audio (WavLM Feature).
\item Mel-spectrograms of the distorted speech (Mel-Spectrogram).
\item Linear spectrograms of the distorted speech (Linear Spectrum).
\end{itemize}
When used as conditional inputs to the LM, all features were projected to the same dimension and padded to the same length to match the dimension of the Enhanced Feature. The results on the Voicefixer GSR test set are presented in Figure \ref{fig5}, where the x-axis represents the SNR of the different features relative to their corresponding clean speech features. The findings indicate that the enhanced features used in our approach exhibit the highest SNR, making them the most suitable choice as conditional inputs for the second-stage LM.

\section{Conclusion}

The OmniGSE framework proposed in this study effectively addresses the challenge of multi-distortion speech enhancement in complex scenarios by combining the strengths of both discriminative and generative approaches. Through a two-stage cross-domain collaborative optimization process, OmniGSE not only performs well in regression tasks such as denoising and dereverberation but also demonstrates strong capabilities in generative tasks like speech restoration. Experimental results demonstrate that OmniGSE outperforms existing methods across multiple benchmarks, with particularly notable improvements in handling compound distortions.

\bibliographystyle{ACM-Reference-Format}
\balance
\bibliography{sample-base}

\end{document}